\documentclass[a4paper,11pt]{article}
\usepackage{jinstpub} % for details on the use of the package, please see the JINST-author-manual
\usepackage{lineno}
\usepackage{siunitx}
\usepackage{xcolor} 
\usepackage{soul}
\usepackage[inkscapelatex=false]{svg}
\usepackage[lofdepth=1]{subfig}

\title{\boldmath Impact of Proton Irradiation on 4H-SiC Low Gain Avalanche Detectors (LGADs)}

\author[a,b]{Y. Satapathy,\note{Corresponding author.}}
\author[a,b]{B.J. Sekely,}
\author[d]{A. Tishelman-Charny,}
\author[c]{T. Yang,}
\author[b]{G. Allion,}
\author[b]{G. Atar,}
\author[b]{P. Barletta,}
\author[c]{C. Haber,}
\author[c]{S.E. Holland,}
\author[b]{J.F. Muth,}
\author[b,1]{S. Pavlidis,}
\author[d]{S. Stucci}

\affiliation[a]{Materials Science and Engineering Department, NC State University\\ Raleigh, NC 27695, USA}
\affiliation[b]{Electrical and Computer Engineering Department, NC State University\\ Raleigh, NC 27695, USA}
\affiliation[c]{Physics Division, Lawrence Berkley National Laboratory\\ Berkeley, CA 94720, USA}
\affiliation[d]{Nuclear and Particle Physics Division, Brookhaven National Laboratory\\ Upton, NY 11973, USA}

\emailAdd{spavlidis@ncsu.edu}

\abstract{Silicon carbide (SiC) particle detectors have the potential to provide time resolutions and robust performance in extreme environments which exceed that of silicon detectors. In this work 4H-SiC low gain avalanche detectors (LGADs) and complementary PiN diodes were irradiated with 2.5 GeV protons at fluences up to 3.33$\times$10$^{14}$ p/cm$^2$. The electrostatic performance of both irradiated and non-irradiated devices was evaluated using current-voltage (I-V) and capacitance-voltage (C-V) measurements. Moreover, charge collection measurements using $\alpha$ particles were also conducted. SiC LGADs displayed a loss in rectification and gain with increasing proton fluence. Additionally, the reduction in capacitance and OFF-state current pointed to compensation of the gain layer as a gain reducing mechanism. The introduction of radiation induced defects also hinders carrier acceleration reducing impact ionization, leading to further gain reduction. However, despite the reduction in device performance, the demonstration of a measurable signal and gain after irradiation points to the potential of SiC LGAD detectors for future high energy physics applications.}


\begin{document}
\maketitle
\flushbottom

\section{Introduction}
\label{sec:intro}

The development of silicon low gain avalanche detectors (Si LGADs) has been driven by the need for ultra-fast solid-state detectors with excellent timing and spatial resolution in high-radiation environments. With their higher timing resolution over traditional Si PN/PiN diodes, Si LGADs are now being adopted in various High-Energy Physics (HEP) applications \cite{fleming_basic_2019}. However, challenges remain in mitigating radiation-induced degradation, which limits their long-term performance under extreme conditions \cite{sadrozinski_4d_2017}.\par

In particular, Si LGADs experience a significant increase in leakage current due to the introduction of radiation induced defects acting as generation sites for charge carriers. Certain irradiation-induced defects in Si LGADs can also compensate active dopants in the gain layer, leading to an overall reduction in gain \cite{feng_study_2022}. To counteract the loss in gain, higher voltage biases need to be applied, which lead to an increased risk of  single-event burnout (SEB) effects \cite{lastovicka-medin_new_2023}. To extend the lifetime of the detector and avoid high leakage currents, bulky, expensive, and complex cooling systems are needed to cool the detectors below 0$^{\circ}$ C  \cite{nagai_cooling_2009}. 

To address these limitations, there is growing interest in using wide bandgap materials such as silicon carbide (SiC) for the fabrication of LGADs. Polytype 4H silicon carbide (4H-SiC) has been widely integrated into power applications, with growth and processing technologies becoming more mature and making this polytype more commercially available. With its large bandgap (3.26 eV) and high threshold displacement energy (C: 21 eV, Si: 35 eV) \cite{barletta2022fast}, SiC-based devices may be more resistant to degradation in harsh radiation environments compared to their silicon counterparts, granting an advantage to a SiC LGAD.

In a hadron collider environment energetic particles crossing the device may interact both through ionizing and non-ionizing processes. The latter case may induce damage due to significant displacement of atoms from their lattice sites. With the expectation of displaying superior radiation hardness, understanding non-ionizing energy loss (NIEL) effects in SiC is of utmost importance. The present work uses 2.5 GeV protons as the irradiation source which may interact with both Coulomb and nuclear forces, effectively causing lattice damage. Similar defects have also been found in neutron irradiated SiC \cite{hazdra_point_2014}. Previous works on proton irradiated SiC materials and devices have displayed compensation in both n-type and p-type 4H-SiC due to generated point defects \cite{kozlovski_model_2015}\cite{emtsev_similarities_2012}, loss of rectification in diodes \cite{rafi_electron_2020}, reduction in charge collection efficiency \cite{he_high-precision_2024}, reduced carrier lifetimes \cite{li_mechanisms_2025}\cite{hazdra_optimization_2018}, and even loss in contact resistivity due to nuclear spallation \cite{kazukauskas_effect_2007}. Radiation induced defects causing these observations in irradiated SiC have been characterized through the use of several techniques including deep level transient spectroscopy (DLTS) \cite{castaldini_deep_2005}, electron paramagnetic resonance (EPR) \cite{lebedev_doping_2000}, photoluminescence (PL) \cite{ruhl_controlled_2018}, and capacitance-voltage profiling \cite{lebedev_doping_2000} in similar studies.

A comprehensive understanding of radiation damage, built on extensive experimental and simulation campaigns, is essential to guide device optimization. This was historically the case for silicon LGADs and is now emerging for SiC. This work provides experimental characterization of SiC LGADs irradiated with 2.5 GeV protons, including current-voltage (I-V), capacitance-voltage (C-V), and alpha-particle charge collection measurements, and lays important groundwork for evaluating their suitability in radiation-intense environments.

\section{Device Fabrication}
\label{sec:fab}

\begin{figure}[htbp]
\centering
\includegraphics[width=1.0\textwidth, page=9]{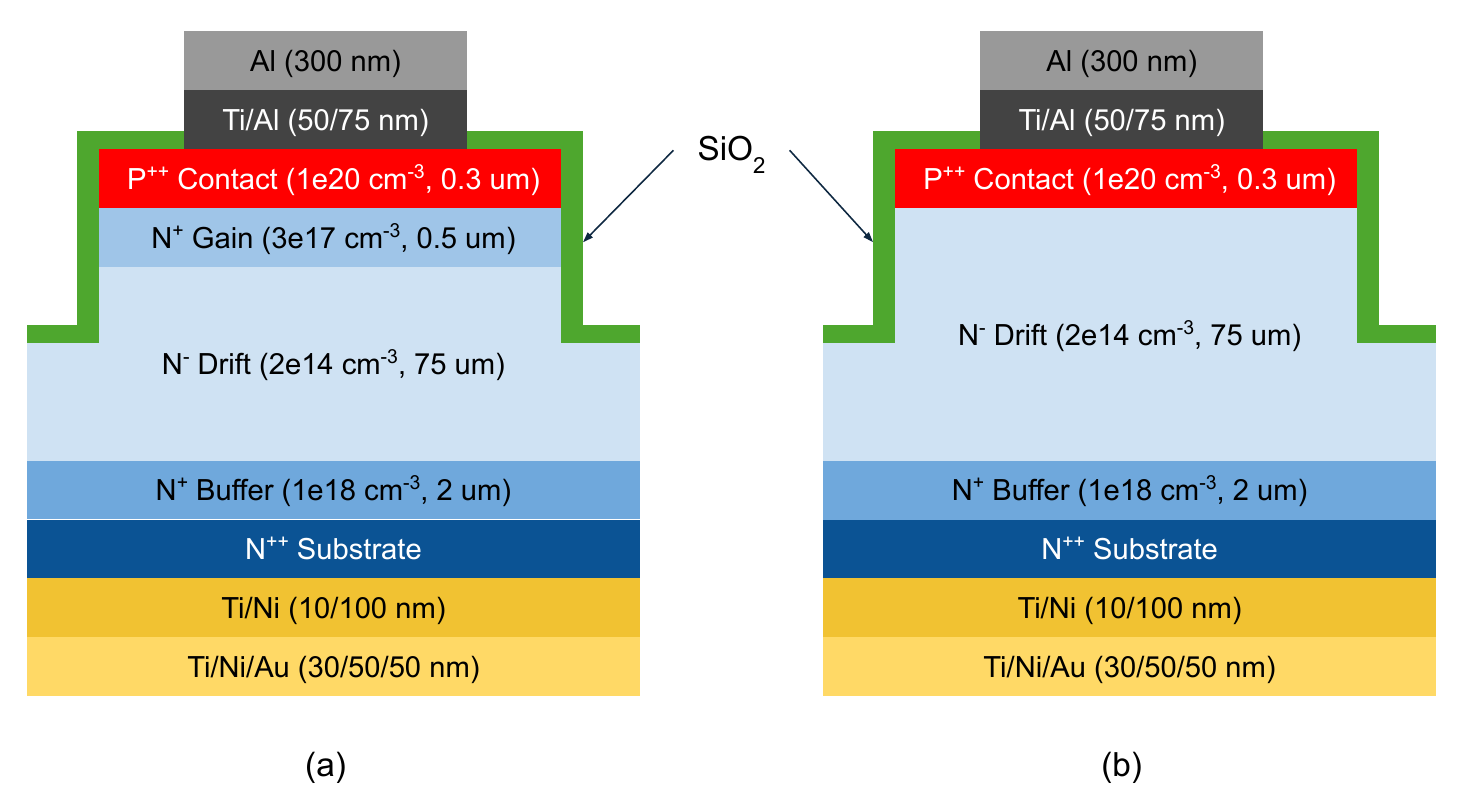}
%\qquad
\caption{Epitaxial stack of a (a) SiC LGAD and (b) SiC PiN diode. The LGAD is identical to the PiN diode with the exception of having an additional moderately doped n-type gain layer. The high electric field in the gain layer promotes charge multiplication through hole-initiated impact ionization.\label{fig:epi_stack}}
\end{figure}

SiC LGADs were fabricated on homoepitaxially grown wafers seen in Figure \ref{fig:epi_stack}. PiN diodes, which have the same epitaxial stack as the LGAD with the exception of the gain layer, were also fabricated alongside LGADs to mitigate variability in device performance due to processing conditions in order to make accurate comparisons.

First, mesa isolation was implemented with fluorine-based reactive ion etching (RIE). An etch depth of 1.3 $\mu$m was measured using a contact profilometer. The samples were subsequently cleaned using solvents and an RCA cleaning procedure. This was followed by dry thermal oxidation to grow a $\sim$50 nm layer of SiO$_2$. Afterwards a thicker 450 nm SiO$_2$ layer was deposited via plasma-enhanced chemical vapor deposition for device passivation. The SiO$_2$ was selectively etched using RIE to expose the SiC surface for metal contact deposition. Ti/Al (50/75 nm) and Ti/Ni (10/100 nm) metal stacks were deposited via electron-beam evaporation on the anode and cathode, respectively. The contacts were annealed at 1000$^{\circ}$C for 5 minutes to form ohmic contacts on the p-type contact layer and n-type substrate. Finally, Al (300 nm) and Ti/Ni/Au (30/50/50 nm) were deposited via electron-beam evaporation on the anode and cathode, respectively, to serve as pad metal for probing during electrical characterization.

\section{Experimental Setup}
\label{sec:setup}

\begin{figure}[htbp]
\centering
\includegraphics[width=.75\textwidth]{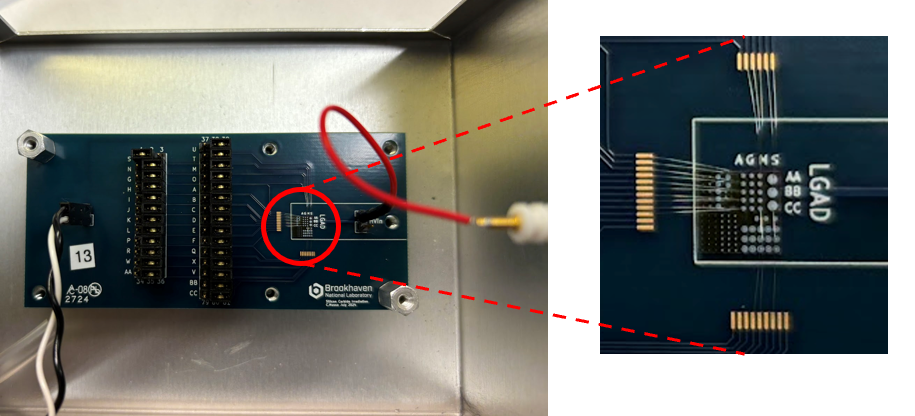}
\caption{SiC LGAD mounted on a PCB board.\label{fig:pcb}}
\end{figure}

\begin{figure}[htbp]
\centering
\includegraphics[width=0.75\textwidth]{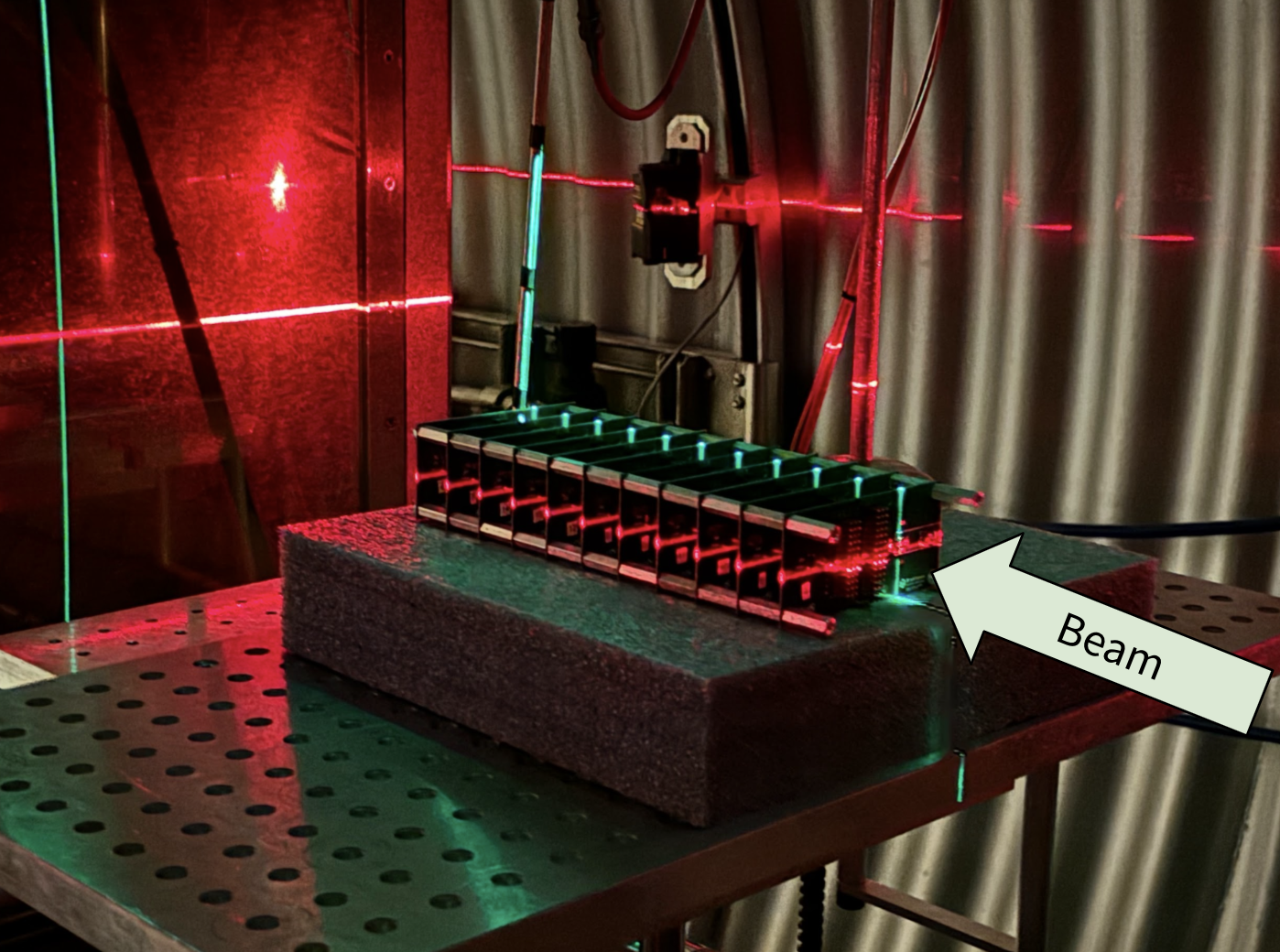}
\caption{A stack of PCB boards aligning all SiC LGADs and SiC PiN diodes exposed to a 2.5 GeV proton beam. The green and red lights are used for alignment.\label{fig:nsrl}}
\end{figure}

The SiC LGADs and PiN diodes were irradiated with 2.5 GeV protons at the NASA Space Radiation Laboratory (NSRL) located at Brookhaven National Laboratory (BNL). Eleven 5$\times$5 mm$^2$ square dies were mounted on printed circuit boards (PCBs) (Figure \ref{fig:pcb}) to facilitate two-probe electrical testing. The PCBs were arranged in a stacked array to ensure uniform exposure of the proton beam across all samples, as shown in Figure \ref{fig:nsrl}. Alignment of the board stackup was performed to ensure the proton beam strikes each device with the same orientation. 

The proton beam, characterized by a Gaussian profile with a full width at half maximum (FWHM) of 2 cm, exposed each die completely during irradiation. Devices were irradiated at three fluences: 1$\times$10$^{13}$ p/cm$^2$, 1$\times$10$^{14}$ p/cm$^2$, and 3.33 $\times$10$^{14}$ p/cm$^2$. Exposing the LGADs to different fluences will give insights on degradation mechanisms of the device as the fluence reaches closer to the expected fluence levels in the High Luminosity Large Hadron Collider (HL-LHC). In future works, we intend to expose devices to larger fluence levels ($\geq$1$\times$10$^{15}$ $n_{eq}$/cm$^2$) in order to more fully explore the expected dose a device might be expected to receive over its lifetime in a future collider.

\begin{figure}[htbp]
\centering
\includegraphics[width=1.0\textwidth,page=7]{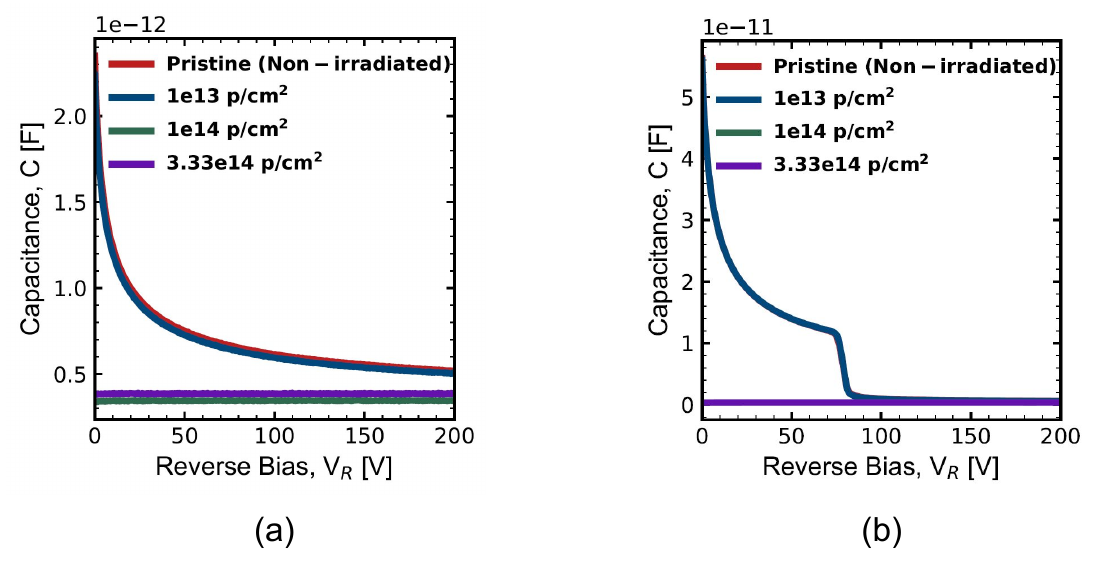}
\caption{C-V characteristics of irradiated and non-irradiated (a) PiN diodes and (b) LGADs. Low proton fluences (1$\times$10$^{13}$ p/cm$^2$) have minimal impact on the capacitance. Higher proton fluences (1$\times$10$^{14}$ p/cm$^2$ and 3.33$\times$10$^{14}$ p/cm$^2$) reduce the capacitance of the devices. Flat capacitance curves indicate compensation in the gain and drift layers. \label{fig:elec_cv}}
\end{figure}

After irradiation, electrical measurements were taken in order to observe any changes in device performance with increasing proton fluence. I-V characteristics (seen in Figures \ref{fig:pin_iv} and \ref{fig:lgad_iv}) were measured using a Keithley 2657A high voltage source measurement unit (SMU). Fluorinert FC-70 was applied to the devices to prevent arcing during measurements. Current compliance values of 10$^{-3}$ A and 10$^{-6}$ A were chosen when measuring the ON-state and OFF-state characteristics respectively. The noise floor of the SMU is $\sim$10$^{-13}$ A. Capacitance-voltage (C-V) characteristics (seen in Figure \ref{fig:elec_cv}) were also measured using a Keithley 4200 semiconductor parameter analyzer. The noise floor of the semiconductor parameter analyzer is $\sim$10$^{-13}$ F. 

\begin{figure}[htbp]
\centering
\includegraphics[width=0.50\textwidth,page=10]{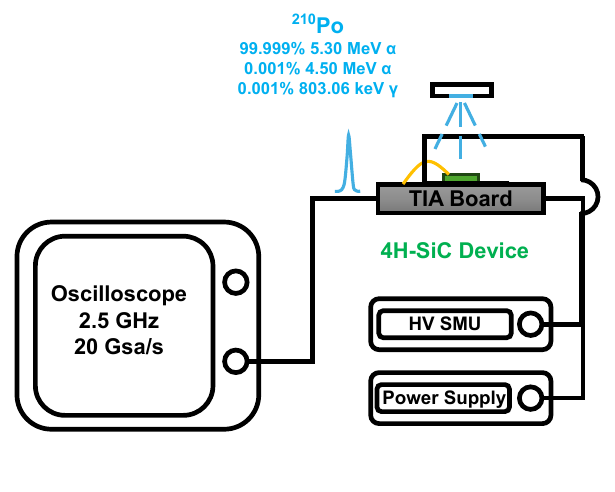}
\caption{Circuit diagram of SiC LGAD/PiN diode connected to a TIA readout board. Keithley 2410 SMU applies a reverse voltage bias of 500 V across the device while the oscilloscope measures the signal generated from the device in response to being exposed to $\alpha$ particles.\label{fig:circuit}}
\end{figure}

To ensure the LGAD is operational after irradiation, the charge collection characteristics were measured. The LGADs and PiN diodes were wire-bonded to a readout board with a trans-impedance amplifier (TIA). The readout board was connected to a Keithley 2410 SMU and an oscilloscope with a 2.5 GHz bandwidth and a sampling rate of 20 GSa/s as seen in Figure \ref{fig:circuit}. The devices were exposed to a $^{210}_{84}$Po $\alpha$-radiation source while reversed biased at 500 V. More information about the setup can be found in \cite{yang_characterization_2024}.

\section{Results}
\label{sec:results}

%This section describes the following results and features seen from measuring the devices before and after irradiation: I-V characteristics, C-V measurements, and charge collection efficiency.

\subsection{Electrostatic Characteristics}

\begin{figure}[htbp]
\centering
\includegraphics[width=1.0\textwidth,page=6]{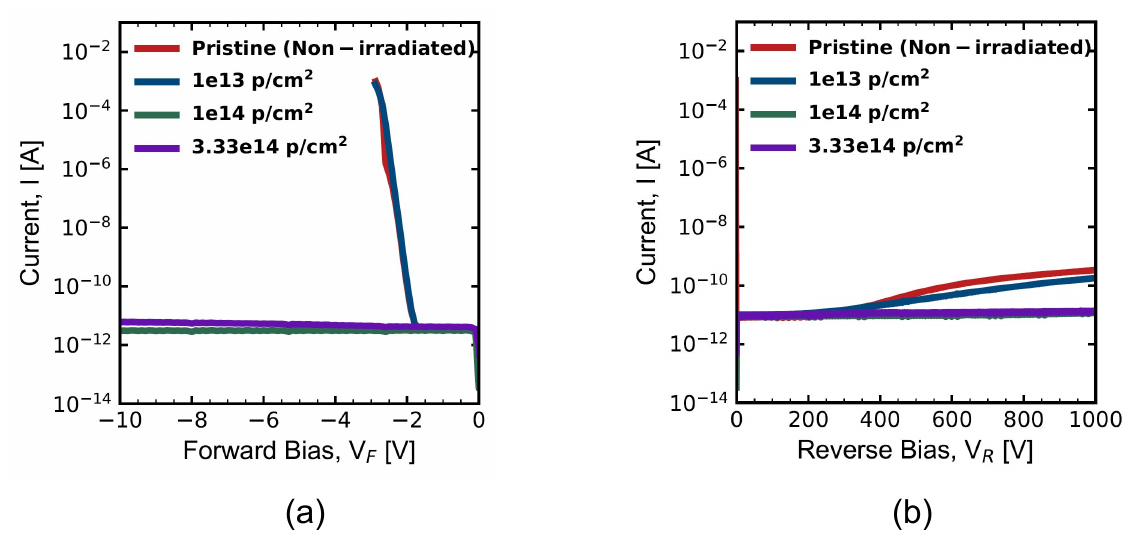}
\caption{(a) ON-state and (b) OFF-state I-V characteristics of irradiated and non-irradiated SiC PiN diodes. A proton fluence of 1$\times$10$^{13}$ p/cm$^2$ shows minimal impact on both the ON-state and OFF-state current. However proton fluences greater than 1$\times$10$^{14}$ p/cm$^2$ significantly reduce both the ON-state and OFF-state current and demonstrate a loss in rectification.\label{fig:pin_iv}}
\end{figure}

\begin{figure}[htbp]
\centering
\includegraphics[width=1.0\textwidth,page=5]{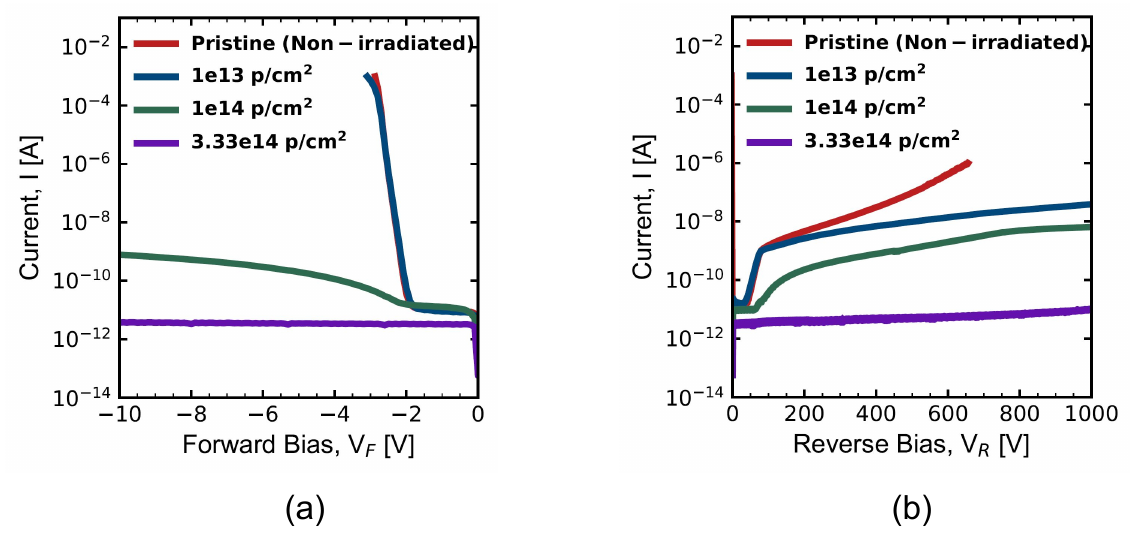}
\caption{(a) ON-state and (b) OFF-state I-V characteristics of irradiated and non-irradiated SiC LGADs. Low proton fluences (1$\times$10$^{13}$ p/cm$^2$) have minimal impact on the ON-state current. All proton fluences significantly reduce the OFF-state current bias. A loss in rectification is observed at proton fluences of 1$\times$10$^{14}$ p/cm$^2$ and greater.\label{fig:lgad_iv}}
\end{figure}

\begin{figure}[htbp]
\centering
\includegraphics[width=1.0\textwidth,page=8]{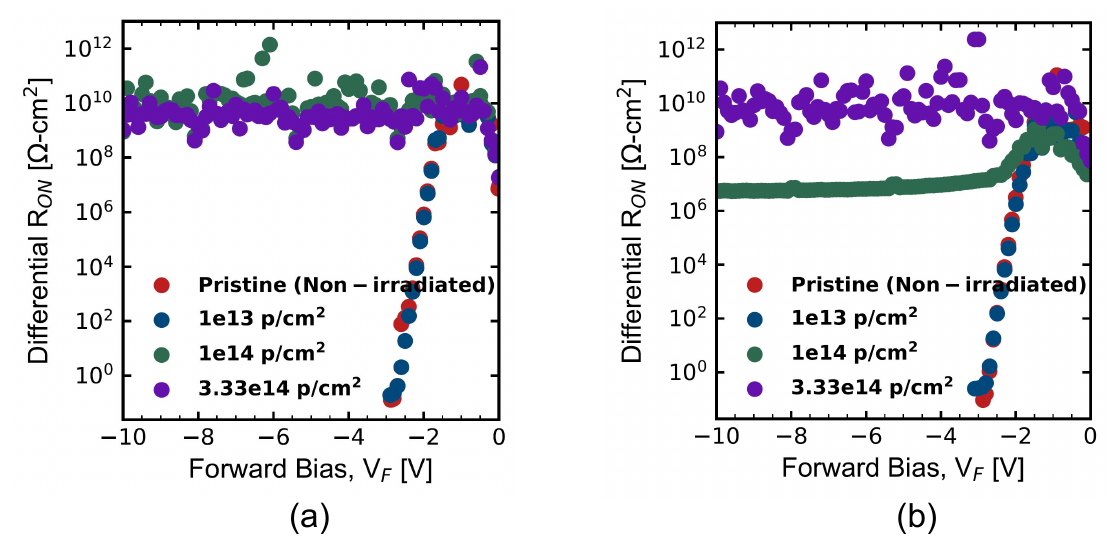}
\caption{Differential ON-state resistance of irradiated and non-irradiated SiC (a) PiN diodes and (b) LGADs. Proton fluence of 1$\times$10$^{13}$ p/cm$^2$s has minimal impact on the differential ON-state resistance. However proton fluences greater than 1$\times$10$^{14}$ p/cm$^2$ significantly increase the differential ON-state resistance.\label{fig:diff_ron}}
\end{figure}

% After irradiation, electrical measurements were taken. I-V characteristics (seen in Figures \ref{fig:pin_iv} and \ref{fig:lgad_iv}) were measured using a Keithley 2657A high voltage source measurement unit (SMU). Fluorinert FC-70 was applied to the devices to prevent arcing. 

According to Figures \ref{fig:pin_iv}, \ref{fig:lgad_iv}, and \ref{fig:diff_ron}, the pristine (non-irradiated) PiN diode and LGAD both exhibited a turn-on voltage (V$_{ON}$) of 2.7 V, rectification ratio (RR) > 10$^7$ (at the voltage where the current reaches compliance), and differential ON-state resistance (R$_{ON}) \leq$0.1 $\Omega$-cm$^2$. Devices irradiated at a fluence of 1$\times$10$^{13}$ p/cm$^2$ do not show significant changes in the ON-state characteristics. The OFF-state (leakage) current starts decreasing when irradiating both the LGAD and PiN diode at a fluence of 1$\times$10$^{13}$ p/cm$^2$. At a fluence of 1$\times$10$^{14}$ p/cm$^2$ and 3.33$\times$10$^{14}$ p/cm$^2$, the ON-state and OFF-state current decreases significantly in both the LGAD and PiN diode. The ON-state current values are comparable to the OFF-state current, displaying a loss in rectification in the device. At higher fluences, the characteristic "step" feature in the OFF-state, which indicates the depletion voltage of the gain layer for the LGADs, disappears. This implies that the gain layer is being electrically deactivated, potentially from a reduction in net doping concentration from compensation to a point where it becomes electrically indistinguishable from the drift layer.\par

The increasing differential ON-state resistance from increasing proton fluence, seen in Figure \ref{fig:diff_ron}, in both the LGAD and PiN diode is hypothesized to be due to reduced carrier lifetime \cite{6907949} and compensation of dopants in the n-type layers, both known to increase ON-state resistance of a semiconductor device. First-order calculations using equations from \cite{baliga_fundamentals_2018} were done to confirm this. Compensation has been shown in literature to be caused by carbon vacancies (V$_C$), silicon vacancies (V$^{+}_{Si}$), and V$_C+$V$_{Si}$ complexes in 4H-SiC \cite{dong_electron_2019} \cite{son_intrinsic_2007}. These defects, also observed in proton irradiated SiC \cite{li_mechanisms_2025}\cite{hazdra_optimization_2018}\cite{karsthof_conversion_2020}, have been seen to be acceptor-like in nature where they trap moving electrons, reducing the net doping levels in n-type SiC (n-SiC). Technology computer-aided design (TCAD) simulations of neutron irradiated SiC PiN diodes presented in \cite{burin_tcad_2024} have shown that models introducing radiation induced defects (Z$_{1/2}$, EH$_{6/7}$, and EH4) display trapping of charge carriers and lead to the same electrostatic characteristics (I-V and C-V measurements) seen in this study. The main defect known to reduce carrier lifetime in is the Z$_{1/2}$ carbon vacancy defect, which has also been shown in literature to increase in concentration with increasing irradiation fluence \cite{kimoto_lifetime-killing_2008}\cite{danno_investigation_2007}. p-type SiC (p-SiC) has also been seen to experience compensation due to irradiation \cite{kozlovski_effect_2015}. Compensation in p-SiC has been observed to be caused by the creation of a Al$_{Si}$-V$_C$ complex \cite{matsuura_mechanisms_2006}. A decrease in the doping level in either the p-SiC contact layer or the n-SiC gain layer will reduce the electric field at the interface and is expected to reduce impact ionization events and ultimately decrease the gain of the LGAD.

The low OFF-state current in the PiN diode is expected due to the large bandgap of SiC. Defects generated deep in the bandgap will not be efficient in generating carriers thermally \cite{rafi_electron_2020}. Compared to SiC, Si LGADs experience an increase in leakage current post-irradiation due to the small bandgap of Si allowing defects to generate more carriers at the same temperature \cite{pellegrini_technology_2014}. Si LGADs also face compensation of dopants in the gain layer through the acceptor removal effect (ARE) where the B atoms in p-type Si are deactivated \cite{feng_study_2022}. This is different from SiC where the N donor atoms in n-SiC gain layer are not deactivated, instead acceptor-like defects are trapping mobile carriers. Therefore in order to improve the radiation hardness of a SiC LGAD, it is essential to fabricate a device that prevents the introduction of these specific defects during irradiation.

The OFF-state current in the pristine LGAD is higher compared to the pristine PiN diode due to the gain mechanism. The decreasing OFF-state current in response to increasing proton fluence in the LGAD is hypothesized to be caused by a reducing electric field at the pn-junction due to compensation in the gain layer. Introducing defects is expected to also act as scattering centers and cause a hindrance in carrier acceleration, which will prevent impact ionization. 

%Capacitance-voltage (C-V) characteristics (seen in Figure \ref{fig:elec_cv}) were measured using a Keithley 4200 semiconductor parameter analyzer. 

In both the LGAD and PiN diode, there is minimal change in the C-V characteristics at a fluence of 1$\times$10$^{13}$ p/cm$^2$. Both devices display a flat C-V curve at higher fluences of 1$\times$10$^{14}$ p/cm$^2$ and 3.3$\times$10$^{14}$ p/cm$^2$. This implies that the drift and gain layers experience compensation to the point where the layers become more intrinsic and also electrically indistinguishable from each other, resulting in a decrease in capacitance. First-order TCAD simulations were done to confirm compensation affecting the C-V characteristics. This phenomena has been seen previously in irradiated SiC PiN diodes \cite{gsponer_neutron_2023}\cite{rafi_electron_2020}, however this is the first time this was observed in a SiC LGAD.  Again, the decrease in doping level in the gain layer is expected to lead to a reduction in gain of the LGAD.

\subsection{Charge Collection Characteristics}

\begin{figure}[htbp]
\centering
\includegraphics[width=1.0\textwidth,page=1]{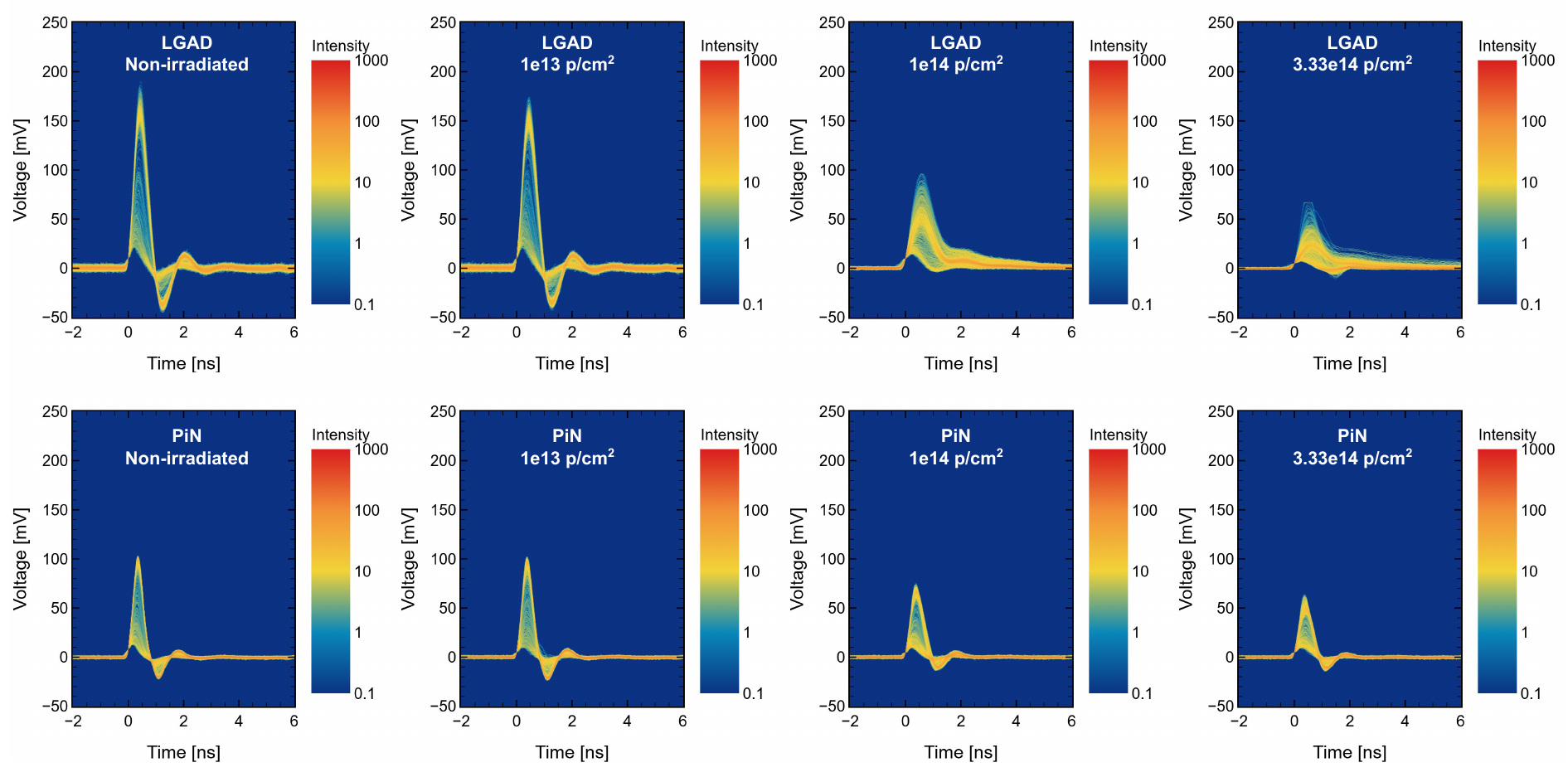}
\caption{Signal pulses intensity distribution of $\alpha$ particles for SiC LGADs and SiC PiN diodes irradiated at various fluences (non-irradiated, 1$\times$10$^{13}$ p/cm$^2$, 1$\times$10$^{14}$ p/cm$^2$, and 3.33$\times$10$^{14}$ p/cm$^2$) at 500 V.\label{fig:signal}}
\end{figure}

 The measured signals, induced by the $^{210}_{84}$Po $\alpha$-radiation source, were recorded from the oscilloscope for each device and is shown in Figure \ref{fig:signal}. The measured signals display a reduction in amplitude for both the LGAD and PiN diode once the proton fluence is greater than 1$\times$10$^{13}$ p/cm$^2$, which clearly indicates a loss of gain in the LGAD. Although at higher fluences both devices showed a loss in rectification, both devices still produced a measurable signal indicating that they are still operational. Similar results were seen in literature where the charge collection efficiency in SiC PiN diodes was reduced when irradiated with a 1.6 GeV proton source at fluences of 3.9$\times$10$^{13}$ p/cm$^2$ and 7.8$\times$10$^{14}$ p/cm$^2$ \cite{he_high-precision_2024}.

\begin{figure}[htbp]
\centering
\includegraphics[width=1.0\textwidth,page=2]{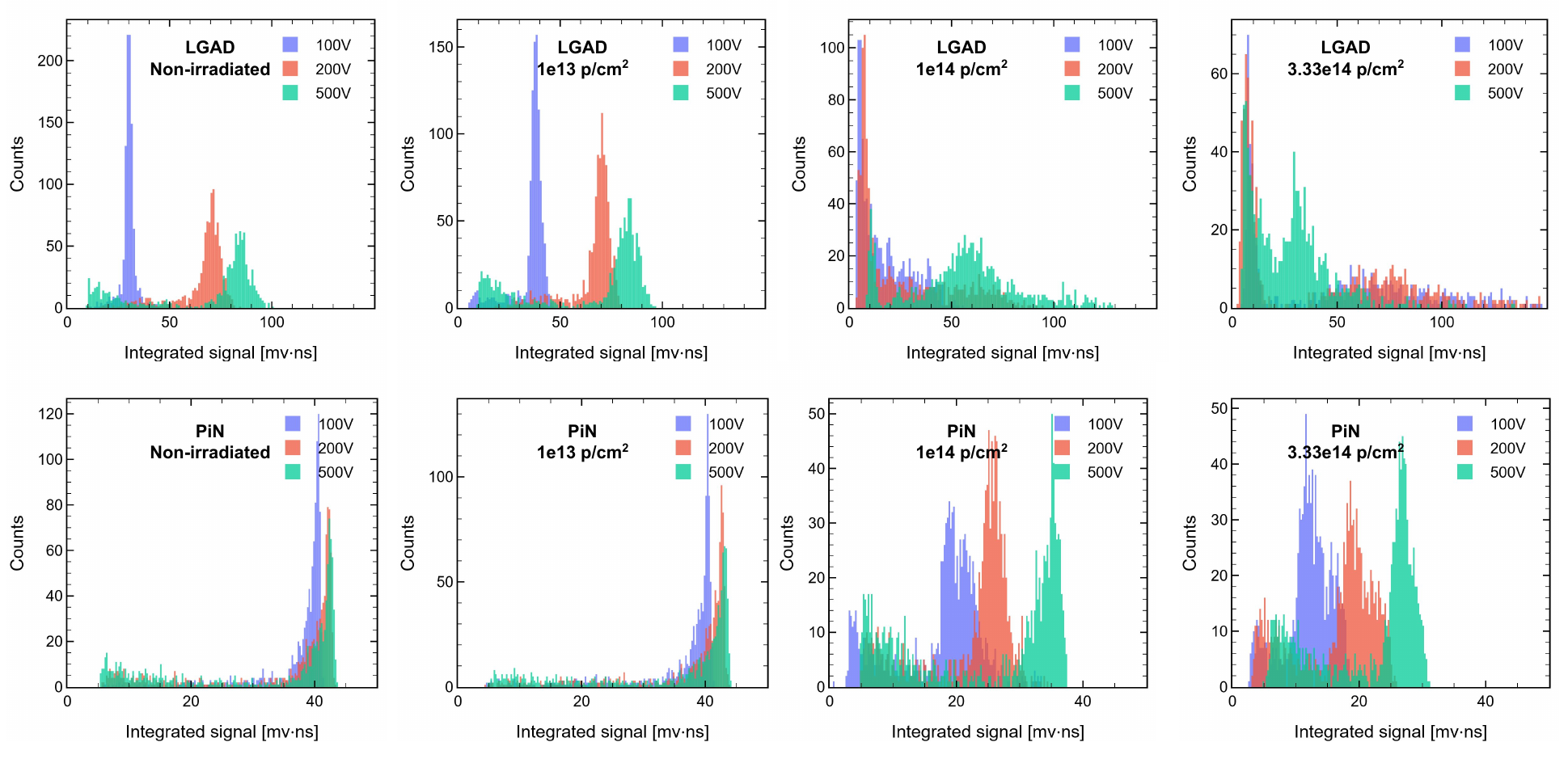}
\caption{Distribution of the integrated signal of SiC LGADs and SiC PiN diodes irradiated at various fluences (non-irradiated, 1$\times$10$^{13}$ p/cm$^2$, 1$\times$10$^{14}$ p/cm$^2$, and 3.33$\times$10$^{14}$ p/cm$^2$) at 100 V (blue), 200 V (red), and 500 V (green) where individual pulses are fitted by gaussian functions.\label{fig:distribution}}
\end{figure}

%ijiwcoij

\begin{figure}[htbp]
\centering
\includegraphics[width=1.0\textwidth,page=3]{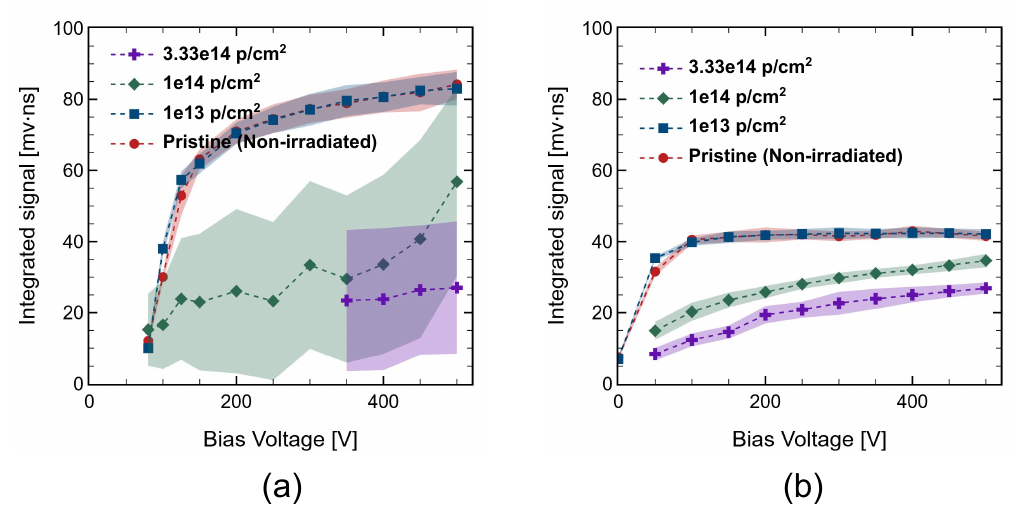}
\caption{Integrated signal vs. bias voltage of the (a) SiC LGAD and (b) PiN diode. The points and shaded regions indicate the amplitude location and standard deviation calculated when fitting the distribution of integrated signals to a gaussian function.\label{fig:intsig}}
\end{figure}

\begin{figure}[htbp]
\centering
\includegraphics[width=0.5\textwidth,page=4]{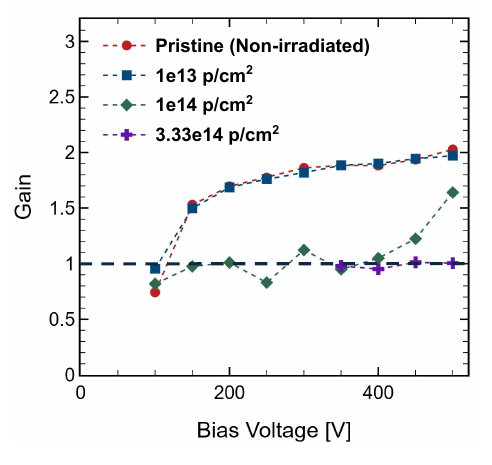}
\caption{Signal gain of the SiC LGAD at various biases after being irradiated at various fluences (non-irradiated, 1$\times$10$^{13}$ p/cm$^2$, 1$\times$10$^{14}$ p/cm$^2$, and 3.33$\times$10$^{14}$ p/cm$^2$).\label{fig:gain}}
\end{figure}

Figure \ref{fig:distribution} shows the distribution of integrated signals of the LGAD and PiN diode at 100 V, 200 V, and 500 V. A gaussian function is used to fit the peaks, afterwards the location of the amplitude and its standard deviation was calculated for each bias voltage and graphed in Figure \ref{fig:intsig}. It was observed that peak broadening and a reduction in amplitude occurs with increasing fluence in the LGADs. The generated defects are not only decreasing the charge collection of the LGAD but also causing larger fluctuations in energy deposited (straggling). The signal gain was calculated by taking the ratio of the integrated signals between the LGAD and PiN diode at each bias voltage and is shown in Figure \ref{fig:gain}. At fluences greater than 1$\times$10$^{13}$ p/cm$^2$, the gain of the LGAD decreases. At the highest fluence there is no observable gain in the LGAD. When exposing the LGAD to a fluence of 1$\times$10$^{14}$ p/cm$^2$ the gain, while reduced at lower bias voltages, was partially replenished when applying a higher bias voltage. 

\section{Conclusion and Outlook}
The presented proton irradiation studies on SiC LGADs bears on their potential for use in high energy physics applications. In comparison to PiN diodes, the LGADs exhibit many of the same changes in electrical performance including a loss in rectification and potential compensation in the epi-layers at higher fluences. However there is also a notable loss in gain with the LGAD. While SiC has a higher threshold displacement energy compared to Si, it still experiences significant changes in its electrical and charge collection characteristics at high fluences. Even at these high fluences, the LGAD is still operational and can approach its original gain value if the bias voltage is increased. For future radiation hard SiC LGADs this study points to the importance of understanding radiation induced defects and their role in reducing gain and charge collection. It is anticipated that further optimization of device designs and understanding these effects may lead to more resilient SiC devices in relevant radiation environments for high energy physics experiments.

\label{sec:conc}

% \appendix
% \section{Some title}
% Please always give a title also for appendices.

\acknowledgments

This material is based upon work supported by the U.S. Department of Energy, Office of Science, Office of High Energy Physics, under Contract Number DE-AC02-05CH11231 and Award Number DE-SC0024252. Additional support was provided by the National Science Foundation under contract No. ECCS-1542015. This work was performed in part at the NCSU Nanofabrication Facility (NNF), a member of the North Carolina Research Triangle Nanotechnology Network (RTNN), which is supported by the National Science Foundation (Grant ECCS-1542015) as part of the National Nanotechnology Coordinated Infrastructure (NNCI). We would also like to thank NSRL for their guidance in helping setup the irradiation experiment.

% Bibliography

%% [A] Recommended: using JHEP.bst file

% \begin{thebibliography}{99}
% \bibliographystyle{JHEP}
% \bibliography{biblio.bib}

% %% or
% %% [B] Manual formatting (see below)
% %% (i) We suggest to always provide author, title and journal data or doi:
% %% in short all the informations that clearly identify a document.
% %% (ii) please avoid comments such as "For a review'', "For some examples",
% %% "and references therein" or move them in the text. In general, please leave only references in the bibliography and move all
% %% accessory text in footnotes.
% %% (iii) Also, please have only one work for each \bibitem.

% % \begin{thebibliography}{99}

% % \bibitem{ATLAS}
% % \textit{Destructive breakdown studies of irradiated LGADs at beam tests for the ATLAS HGTD},
% % L.A. Beresford et al 2023 JINST 18 P07030

% % \bibitem{a}
% % Author,
% % \emph{Title},
% % \emph{J. Abbrev.} {\bf vol} (year) pg.

% % \bibitem{b}
% % Author,
% % \emph{Title},
% % arxiv:1234.5678.

% % \bibitem{c}
% % Author,
% % \emph{Title},
% % Publisher (year).

% \end{thebibliography}

\providecommand{\href}[2]{#2}\begingroup\raggedright\endgroup

\end{document}